\documentclass[aps,prb,reprint,nofootinbib,twocolumn,superscriptaddress,showpacs,showkeys,longbibliography,amsmath,amssymb]{revtex4-2}
\usepackage[utf8]{inputenc}
\usepackage{graphicx}
\usepackage{dcolumn}
\usepackage{bm}
\usepackage{braket}
\usepackage{subfigure}
\usepackage[english]{babel}
\usepackage{times} 
\usepackage{helvet}
\usepackage{courier}
\usepackage{amsmath} 
\usepackage[colorlinks=true, linkcolor=blue]{hyperref}

\begin{document}
\preprint{APS/123-QED}
\title{Non-Hermitian Mosaic Maryland model}

\author{Zhenning Wang}
\affiliation{Department of Physics, Jiangsu University, Zhenjiang, 212013, China}
\author{Ni Lu}
\affiliation{Department of Physics, Jiangsu University, Zhenjiang, 212013, China}
\author{Dan Liu}
\affiliation{Department of Physics, Jiangsu University, Zhenjiang, 212013, China}
\author{Xiaosen Yang}\email {yangxs@ujs.edu.cn}
\affiliation{Department of Physics, Jiangsu University, Zhenjiang, 212013, China}
\author{Xianqi Tong}\email {xqtong@ujs.edu.cn}
\affiliation{Department of Physics, Jiangsu University, Zhenjiang, 212013, China}

\date{\today}




\begin{abstract}
We introduce the non-Hermitian mosaic Maryland model, where a discrete modulation period and a non-Hermitian phase are incorporated into the potential, rendering the originally exactly solvable system generally non-integrable. This model provides a unique platform to investigate how structural modulation governs localization in complex quasiperiodic potentials. Using Avila’s global theory, we analytically derive the exact Lyapunov exponent and obtain explicit formulas for the complex mobility edges. Remarkably, for modulation periods $\kappa \ge 2$, the system intrinsically hosts $\kappa-1$ robust extended bands that persist independently of the potential strength and non-Hermiticity. We further characterize the topological nature of these phases via the spectral winding number. Unlike the standard Maryland model, the mosaic modulation induces mobility edges, and the resulting phase transitions are continuous, reflecting the non-integrable nature of the system. Numerical calculations of the inverse participation ratio and fractal dimension confirm the analytical predictions for the asymptotic form of the mobility edges in the large non-Hermiticity limit. This work establishes structural design as a powerful degree of freedom for engineering wave transport and enhancing the robustness of extended states in non-Hermitian systems.
\end{abstract}

\maketitle

\section{INTRODUCTION}
The study of non-Hermitian (NH) physics \cite{79,49} has fundamentally reshaped our understanding of quantum phases and wave transport. Unlike their Hermitian counterparts, NH systems exhibit a plethora of exotic phenomena, such as the NH skin effect \cite{81,68,76,74,73,70,69,77,78,104,105,106,114}, breaking of the parity-time symmetry \cite{52,44} and the breakdown of the conventional bulk-boundary correspondence \cite{55,106,111,67,55,112,113,115}. In particular, the interplay between non-Hermiticity and quasiperiodic order has emerged as a fertile ground for discovering unique localization-delocalization transitions and mobility edges \cite{53,63,60,88,89,87,117} in the complex energy plane. While early research primarily focused on random disorder \cite{2,5,17,92}, recent attention has shifted toward deterministic quasiperiodic potentials \cite{53,83}, where the competition between ergodic and localized phases can be rigorously analyzed.

In the landscape of one-dimensional quasiperiodic systems \cite{101,102,103,7}, two models stand out as paradigms: the Aubry-Andr\'{e}-Harper (AAH) model \cite{14,102,103,116} and the Maryland model \cite{1}. The AAH model, characterized by a bounded cosine potential, is renowned for its self-duality and the existence of a metal-insulator transition at a finite potential strength \cite{30,61,118}. Moreover, there are energy-related localized transitions in low-dimensional quasiperiodic systems \cite{119}.The concept of mobility edge (ME) \cite{93} introduces a critical energy to separate the coexisting extended states and localized single-particle eigenstates under the same set of parameters \cite{57}. Notably, the standard AAH model does not include ME because the standard AAH model exhibits self-duality at the transition point between lattice space and momentum space transformations. And various generalized AAH models have already been developed and their ME can be solved analytically, which demonstrates that they can achieve localized transitions with ME \cite{21,23,24,27,41,51,52,53,58,90,40}.

Distinct from this, the Maryland model features an unbounded tangent-type potential and is celebrated as one of the few exactly solvable quasiperiodic models, owing to its integrability via a mapping to a dynamical Floquet problem \cite{42,32,31,28,40}. Recently, NH extensions of these models have been intensively explored, revealing that non-Hermiticity can induce rich spectral topologies and modify the universality classes of localization transitions \cite{13,11,54,72,81,82,85,80,75,70,66,9,8,34,22,51}. However, most existing studies remain confined to homogeneous quasiperiodic potentials \cite{45}, leaving the effects of discrete structural modulation largely unexplored.

The concept of mosaic lattices introduces a geometric degree of freedom to this problem. In a mosaic model, the quasiperiodic potential is applied selectively at sites spaced by a discrete period $\kappa$, rather than at every site. This periodic interruption of the potential creates distinct ballistic channels within the lattice, offering a mechanism to engineer transport properties through geometry itself, complementary to tuning the potential  \cite{59,64}. While mosaic modulation has been studied in the context of bounded AAH potentials \cite{65,59}, its impact on singular, unbounded potentials, such as that of the Maryland model \cite{39,107,108,109,110}, remains an open question. The combination of the Maryland model's analytical tractability with the symmetry-breaking nature of mosaic modulation promises to unveil novel localization mechanisms.

\begin{figure*}
    \centering
    \includegraphics[width=1\linewidth]{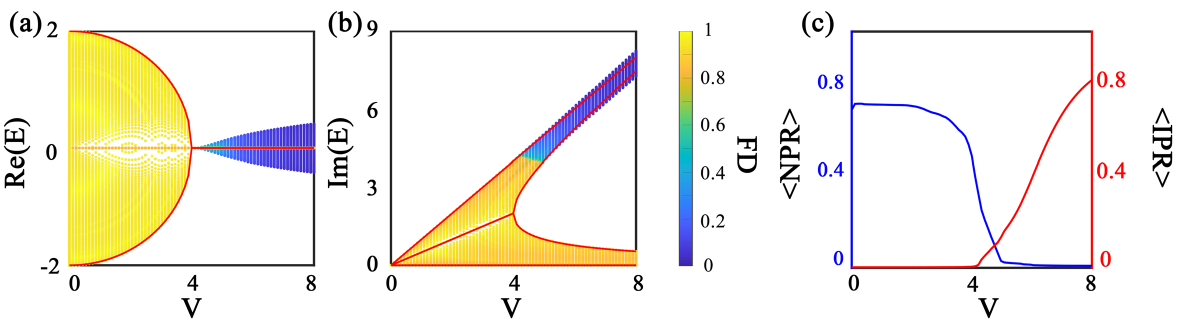}
    \caption{Localization landscapes and global phase transitions of the non-Hermitian mosaic Maryland model for $\kappa=2$. (a) Real part of the energy spectrum Re(E) as a function of the potential strength $V$. The color scale encodes the FD. The red solid lines correspond to the analytical mobility edges derived in Eq. (\ref{eq:ME_kappa2}). (b) Imaginary part of the energy spectrum Im(E) versus $V$. The red lines tracking the phase boundaries analytically match Eq. (\ref{eq:ME_kappa2}). (c) Evolution of global localization indices: $\langle \text{IPR} \rangle$, red solid line, right axis and $\langle \text{NPR} \rangle$, blue solid line, left axis. The kinks clearly demarcate the transition from the fully extended phase to the mobility edge phase. System parameters are set as $L=610$ and $\epsilon=1.8$.}
    \label{fig:placeholder1}
\end{figure*}
In this work, we synthesize these concepts by introducing and rigorously analyzing the non-Hermitian mosaic Maryland model. By imposing the singular Maryland potential onto a mosaic lattice with spacing $\kappa$, we establish a rich theoretical platform that transitions from being exactly solvable ($\kappa=1$) to generally non-integrable ($\kappa \ge 2$). Despite this loss of integrability, we analytically derive the exact Lyapunov exponent $\Gamma(E)$ using Avila's global theory and explicitly determine the complex mobility edges. Crucially, our analysis reveals a rigid structural topological constraint: for any $\kappa \ge 2$, the system intrinsically hosts $\kappa-1$ robust extended bands. These bands persist independently of the potential strength $V$ or the non-Hermiticity parameter $\epsilon$, a mechanism fundamentally distinct from conventional energy-dependent mobility edges. 

The remainder of this paper is organized as follows. In Sec. \ref{model and analytic theory}, we formally define the model and construct the analytical framework. Section \ref{Case Study} provides a detailed case study of the $\kappa=2$ modulation, mapping out the mobility edges and demonstrating the nature of the phase transitions. In Sec. \ref{LEVEL SPACING STATISTICS AND SPECTRAL CORRELATIONS}, we investigate the complex level spacing statistics to rigorously demonstrate the integrability breaking induced by the discrete mosaic modulation. In Sec. \ref{Topological Characterization via Winding Number}, we characterize the topological signatures of these phases via the spectral winding number. Followed by our conclusions in Sec. \ref{Conclusions}.

\section{Model and Analytical Theory}
\label{model and analytic theory}
We consider a non-perturbative NH extension of the Maryland model defined on a one-dimensional tight-binding lattice:
\begin{equation}
\psi_{n+1} + \psi_{n-1} + V_n \psi_n = E\psi_n,
\end{equation}
where $\psi_n$ is the wave function amplitude at site $n$ and the hopping energy is set to unity. The mosaic-type potential $V_n$ is introduced by selectively applying the incommensurate modulation at every $\kappa$-th site:
\begin{equation}
V_n = V \delta_{n, \kappa m} \tan(\pi \alpha n + \theta + i\epsilon), \quad m \in \mathbb{Z},
\end{equation}
where $V$ and $\theta$ represent the potential strength and the global phase, respectively. The parameter $\kappa$ denotes the mosaic spacing, and $i\epsilon$ introduces non-Hermiticity into the complex potential \cite{96}. The irrational parameter $\alpha$ is chosen to have typical Diophantine properties with a zero irrationality measure $L(\alpha)=0$, such as the inverse golden ratio $\alpha = (\sqrt{5}-1)/2$. For $\kappa=1$, the Hamiltonian recovers the homogeneous non-Hermitian Maryland model. For $\kappa \geq 2$, the potential is interrupted periodically, creating propagation channels that fundamentally alter the localization-delocalization transitions.

To analytically determine the localization properties, we employ Avila's global theory to calculate  $\Gamma(E)$. According to this theory, the behavior of $\Gamma(E)$ for generic $\epsilon$ is rigorously determined by its asymptotic form in the large NH limit ($\epsilon \rightarrow \infty$). In this asymptotic regime, the singular tangent potential behaves as $\tan(\pi\alpha n + \theta + i\epsilon) \rightarrow i$, effectively shifting the real energy by a pure imaginary potential $-iV$, see Appendix \ref{Appendix A}. Consequently, the spectral function $f$ is derived from the unit cell transfer matrix, yielding:
\begin{equation}\label{eq:Gamma}
\Gamma(E) = \frac{1}{\kappa} \max \{ \ln(2f), 0 \},
\end{equation}
where the spectral function $f$ is derived from the unit-cell transfer matrix:
\begin{equation}
f = \max_{\pm} \left| \frac{\chi a_{\kappa} - 2a_{\kappa-1} \pm G}{4} \right|.
\end{equation}
Here, motivated by the asymptotic limit, we define the effective complex energy parameter $\chi = E - iV$, and the discriminant $G$ as:
\begin{equation}
G = \sqrt{\chi^2 a_{\kappa}^2 - 4\chi a_{\kappa} a_{\kappa-1} + 4a_{\kappa} a_{\kappa-2}},
\end{equation}
with the polynomial terms $a_\kappa$ given by:
\begin{equation}
a_\kappa = \frac{1}{\sqrt{E^2-4}} \left[ \left( \frac{E+\sqrt{E^2-4}}{2} \right)^\kappa - \left( \frac{E-\sqrt{E^2-4}}{2} \right)^\kappa \right].
\end{equation}
The complex MEs are identified by the condition $\Gamma(E) = 0$, which corresponds to $2f=1$. In the following sections, we examine the $\kappa=2$ case in detail to illustrate the emergence of mobility edges in the complex energy plane.
\begin{figure*}
    \centering
    \includegraphics[width=1\linewidth]{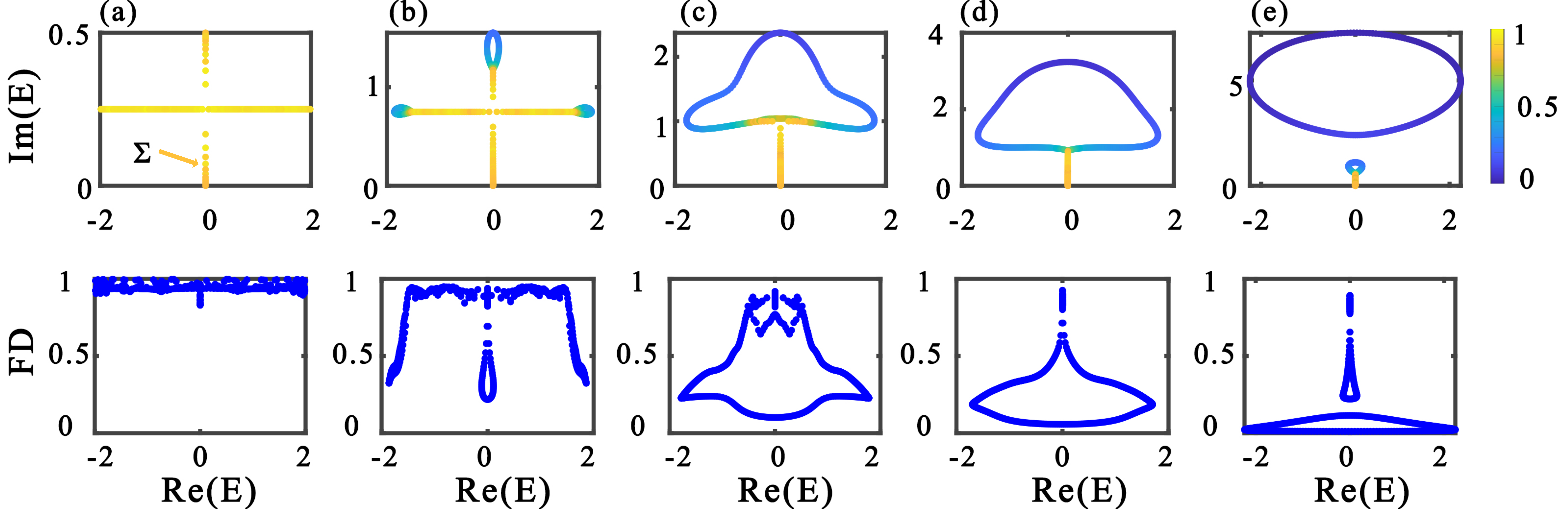}
    \caption{Spectral evolution and localization characteristics of the non-Hermitian mosaic Maryland model. The upper panels display the complex energy spectrum, with extended states ($\Sigma$) shown in yellow and localized states in blue, while the lower panels show the FD of the corresponding eigenstates $\psi_n$. The system is simulated on a lattice of size $L=610$ under periodic boundary conditions with fixed parameters $\kappa=2$,  $\epsilon=0.8$,  $\theta=0$ and $\alpha = 377/610$. The subplots correspond to increasing potential strength: (a) $V=0.5$ (extended phase) and (b) $V=1.5$. (c) $V=2$ (Three loops merged into one) and (d) $V=2.5$ the localized phase increases; (e) $V=5$ (almost fully localized phase, except at $E=0$). }
    \label{fig: placeholder2}
\end{figure*}

\section{Case Study: $\kappa=2$}
\label{Case Study}

We first focus on the representative case of $\kappa=2$. By substituting the lower-order spectral polynomials $a_1=1$ and $a_2=E$ into the general framework of Appendix \ref{Appendix A}, we obtain the explicit analytical condition for the ME:
\begin{equation}
\left | \chi E-2 \right | = 2,
\label{eq:ME_kappa2}
\end{equation}
where $\chi = E - iV$.

In Figs.\ref{fig:placeholder1}(a) and (b), we present the evolution of the energy spectrum with respect to $V$. The localization nature of the eigenstates is characterized by the FD, defined as \cite{98,99}:
\begin{equation}
D_2 = - \lim_{L \to \infty} \frac{\ln\text{IPR}_n}{\ln L},
\end{equation}
where the IPR for the $n$-th eigenstate is given by \cite{100}:
\begin{equation}
\text{IPR}_n = \frac{\sum_{j=1}^{L} |\psi_{j}^n|^4}{\left( \sum_{j=1}^{L} |\psi_{j}^n|^2 \right)^2}.
\end{equation}

In the phase diagram, blue regions ($D_2 \to 0$) correspond to localized states, while yellow regions ($D_2 \to 1$) indicate extended states. The boundaries between these regions delineate the critical $V$ values for the phase transitions.

To provide a quantitative characterization of the global phase diagram, we compute the mean values of the localization indices over the relevant spectral range: the average IPR, denoted as $\langle \text{IPR} \rangle$, and the average Normalized Participation Ratio $\langle \text{NPR} \rangle$, defined as:
\begin{equation}
\langle \text{NPR} \rangle = \frac{1}{L} \sum_{n}^{L} \left( L \times \text{IPR}_n \right)^{-1},
\end{equation}
where $L$ is the number of eigenstates included in the average. Physically, $\langle \text{IPR} \rangle \sim \mathcal{O}(1)$ signals localization, whereas $\langle \text{NPR} \rangle \to 1$ indicates extended states.

As shown in Fig.~\ref{fig:placeholder1}(c), the system exhibits distinct behaviors in different regimes. While the smooth variation of the curves reflects the continuous nature of the ME sweeping through the spectrum, the two sharp kinks clearly identify the phase boundaries. A sharp decline in $\langle \text{NPR} \rangle$ accompanied by the onset of a non-zero $\langle \text{IPR} \rangle$ marks the transition from the fully extended phase to the ME phase. Conversely, the gradual increase of $\langle \text{IPR} \rangle$ towards unity signals the progressive entry into the localized phase. Combining the spectral map in Fig.~\ref{fig:placeholder1}(a) and the global indices in Fig.~\ref{fig:placeholder1}(c), we determine the critical point for the $\kappa = 2$ case to be approximately $V_1\approx 4.2$.

The evolution of the spectral phases is depicted in Fig.~\ref{fig: placeholder2}. In the fully extended phase ($V < V_1$), all eigenstates become delocalized. The energy spectrum condenses into a continuous set $\Sigma$, which converges to the ''cross-shaped'' structure shown in Fig.~\ref{fig: placeholder2}(a). This implies that $\Gamma(E) = 0$ for all energies on this locus [Note: Here the symbol $\Gamma$ denotes both the set and the  $\Gamma(E)$; context distinguishes them, where a zero  $\Gamma(E)$ typically indicates extension].

In the intermediate ME phase ($V > V_1$), the system exhibits a coexistence of localized and extended states, see Figs.~\ref{fig: placeholder2}(b)-(e). The spectrum bifurcates into distinct topological regions, where yellow areas correspond to extended eigenstates and blue areas indicate localized ones. The spectrum associated with localized states forms a discrete set of points in the complex plane, indicating strong confinement. Geometrically, the three endpoints corresponding to the cross gradually expand from shrunk points into loops. The upper loop emerges first; as $V$ increases, the lateral loops subsequently appear. During the continued increase of $V$, the three loops gradually merge into a shell devoid of internal points. Subsequently, this shell splits into two loops: a large loop located at the top that exists independently, and a small loop situated at the bottom that connects with the extended state region.

We emphasize that the observed phase transition behavior and the specific critical value $V_1$ are intrinsic to the $\kappa = 2$ mosaic structure. The phase diagram depends strongly on the modulation period $\kappa$, as the geometry of the ballistic channels fundamentally alters the localization landscape. Furthermore, the topological properties of the energy spectrum are jointly governed by $\kappa$, $V$, and $\epsilon$, providing an additional dimension for characterizing these non-Hermitian phase transitions. A detailed discussion and the analytical derivations of the mobility edges for higher-order modulations, specifically $\kappa = 3$ and $4$, are provided in Appendix~\ref{Appendix B}.

\section{LEVEL SPACING STATISTICS AND SPECTRAL CORRELATIONS}
\label{LEVEL SPACING STATISTICS AND SPECTRAL CORRELATIONS}

\begin{figure}[htbp]
    \centering
    \includegraphics[width=\linewidth]{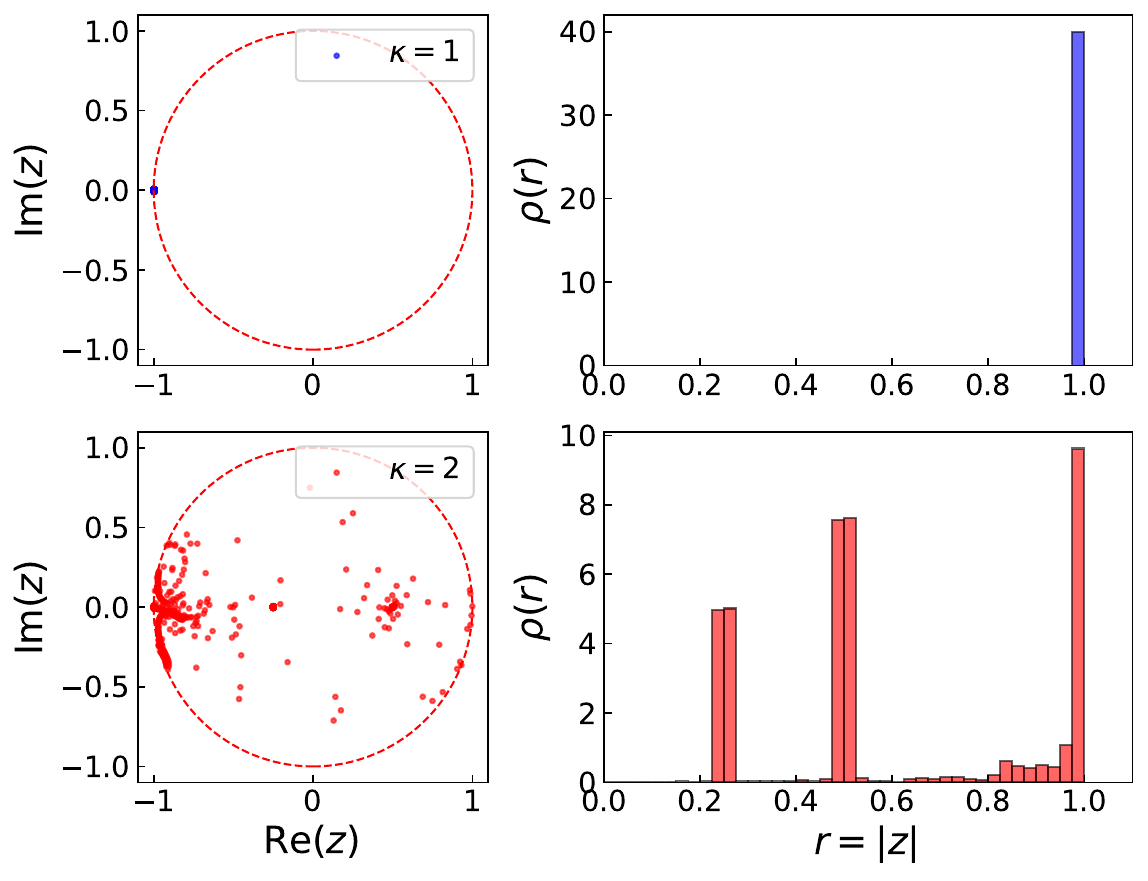} 
    \caption{Complex level spacing statistics for the non-Hermitian Maryland model with different modulation periods. The scatter plots (left column) display the complex spacing ratio $z_i$ in the complex plane, while the histograms (right column) show the probability density $\rho(r)$ of its modulus $r = |z_i|$ for the bulk eigenstates. (a, b) For the standard model ($\kappa=1$), the rigid one-dimensional loop structure of the spectrum forces $z_i$ to strictly cluster around $-1$ on the unit circle (red dashed line), yielding a highly peaked geometric distribution with $\langle r \rangle \approx 0.998$. (c, d) For the mosaic model ($\kappa=2$), The eigenvalues deviate from the 1D trajectories and scatter into the complex plane. However, rather than forming a featureless uniform 2D distribution, $\rho(r)$ exhibits distinct discrete oscillations (multiple peaks), and the mean ratio drops to $\langle r \rangle \approx 0.598$. System parameters are set as $L=1597$, $V=4.0$, and $\epsilon=0.5$.}
    \label{fig_level_spacing}
\end{figure}

To rigorously substantiate the mosaic-induced transition from an exactly solvable 1D limit to a more complex spectral regime, we investigate the complex level spacing statistics of the Hamiltonian. In non-Hermitian systems, the conventional real-energy level spacing is generalized to the complex plane by analyzing the nearest-neighbor (NN) and next-nearest-neighbor (NNN) distances. For each complex eigenvalue $E_i$, we define the complex spacing ratio as 
\begin{equation}
    z_i = \frac{E_i^{NN} - E_i}{E_i^{NNN} - E_i},
\end{equation} 
where $E_i^{NN}$ and $E_i^{NNN}$ are the closest and second-closest eigenvalues to $E_i$ in the Euclidean metric, respectively. We then examine the mean modulus of this ratio, $\langle r \rangle = \langle |z_i| \rangle$, alongside the distribution of $z_i$ in the complex plane and the probability density $\rho(r)$. To avoid statistical artifacts, eigenvalues corresponding to singularity-protected persistent extended bands on the real axis are rigorously excluded from the ensemble.

The statistical results, depicted in Fig.~\ref{fig_level_spacing}, reveal a profound geometrically driven physical transition. For the standard non-Hermitian Maryland model ($\kappa=1$), its point spectrum is subject to a rigid geometric constraint: the eigenvalues collapse onto a one-dimensional analytical closed loop embedded in the complex plane \cite{1}. Because the eigenvalues are confined to a 1D curve, the two nearest neighbors for any bulk state predominantly lie on opposite sides along the trajectory with locally equal spacing, forcing $|z_i| \approx 1$. Consequently, the distribution of $z_i$ tightly clusters on the unit circle near $\text{Re}(z)=-1$, yielding a sharply peaked density $\rho(r)$ with $\langle r \rangle \rightarrow 1$ [see Figs.~\ref{fig_level_spacing}(a) and (b)]. 

In fundamentally sharp contrast, introducing the mosaic modulation ($\kappa = 2$) actively destroys this strict 1D analytical integrability. As shown in Fig.~\ref{fig_level_spacing}(c), the macroscopic 1D spectral loop dissolves, releasing the eigenvalues to scatter into the two-dimensional complex plane. However, the corresponding histogram $\rho(r)$ [Fig.~\ref{fig_level_spacing}(d)] does not transition into a smooth Wigner-like surmise of the Ginibre unitary ensemble (GinUE, $\langle r \rangle \approx 0.738$), nor does it follow a featureless 2D Poisson distribution ($\langle r \rangle \approx 0.667$) expected for uncorrelated 2D spectra. Instead, $\rho(r)$ displays pronounced discrete oscillations (multiple peaks), and the mean spacing ratio stabilizes at $\langle r \rangle \approx 0.598$. 

It is crucial to note that $\langle r \rangle \approx 0.598$ being lower than the 2D Poisson limit signifies the presence of strong local level clustering rather than chaotic level repulsion. This clustering, along with the distinct oscillatory features in $\rho(r)$, is a statistical hallmark of the highly structured, possibly fractal or multi-banded substructures residing within the complex spectrum—a characteristic feature of non-Hermitian quasiperiodic mixed phases. Therefore, while the mosaic modulation unambiguously drives the system out of the exactly solvable 1D limit, it does not lead to fully ergodic quantum chaos. Rather, it gives rise to a intermediate phase with non-standard geometric spectral correlations that defy conventional random matrix universality classes.

\section{Topological Characterization via Winding Number}
\label{Topological Characterization via Winding Number}

\begin{figure}[htbp]    
\centering    
\includegraphics[width=1\linewidth]{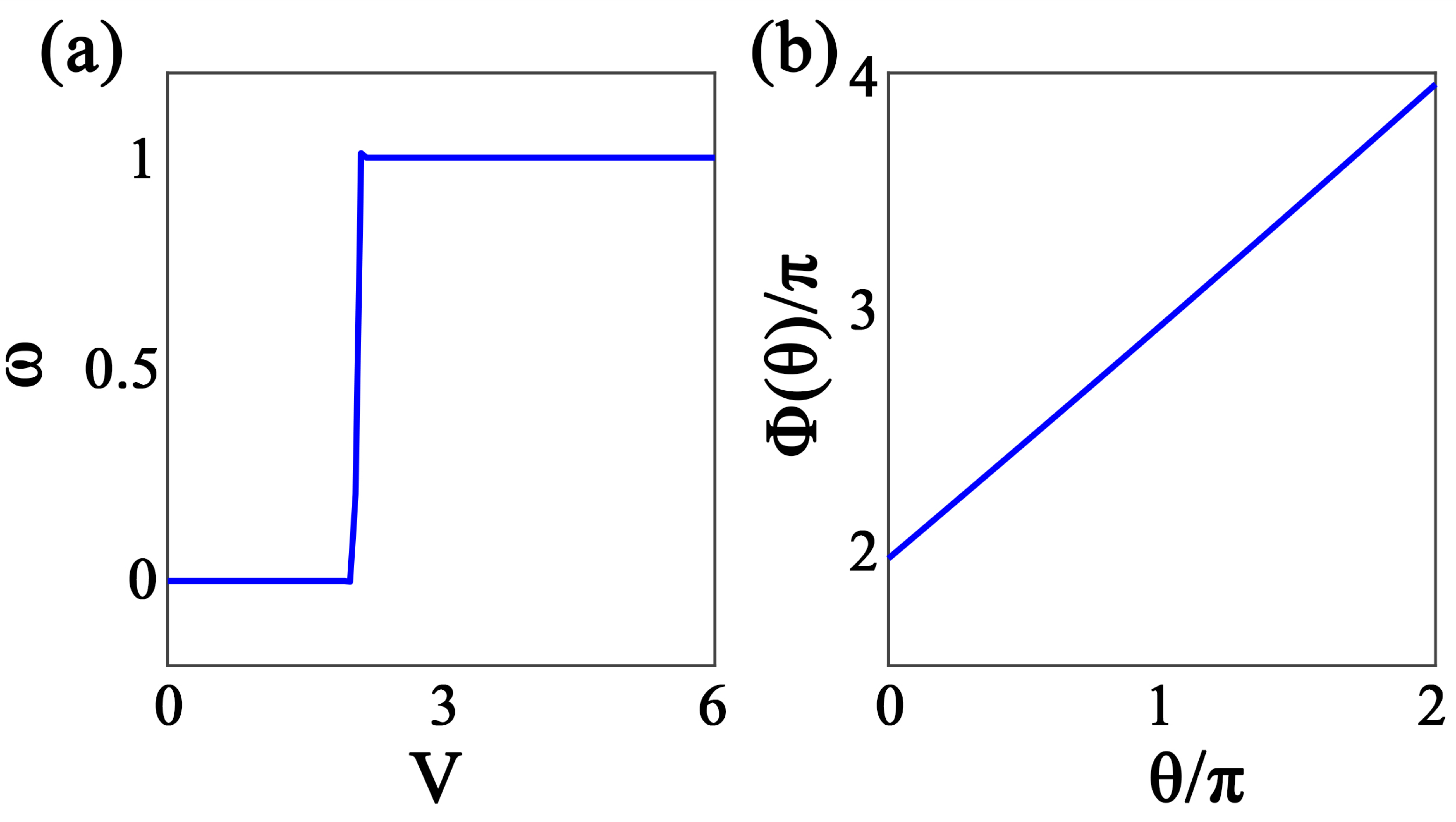}  
\caption{Topological characterization via winding number for $\kappa=2$. (a) Evolution of the winding numbers for $E_{B} = 1.27 + i(V-0.6) $  as a function of $V$. Parameters: $\epsilon = 0.8$, $L = 610$. (b) Geometric phase accumulation for $E_{B}$ as $\theta$ varies from $0$ to $\pi$, calculated at $V = 2.1$ . The linear increase for $E_{B}$ indicates a non-trivial winding topology ($\omega=1$).}
\label{fig:placeholder3}
\end{figure}
To characterize the topological nature of the phase transitions, we introduce a spectral winding number defined under periodic boundary conditions \cite{50,43}. By imposing a phase $\theta$ across the system boundary—physically equivalent to threading a magnetic flux through a ring geometry—we probe the sensitivity of the complex spectrum to the boundary conditions. For a chosen reference base energy $E_B$, the winding number $\omega(E_B)$ is defined as \cite{62,50,95}:
\begin{equation}
    \omega(E_B) = \lim_{L \to \infty} \frac{\kappa}{2\pi i} \int_{0}^{\pi} d\theta \frac{\partial}{\partial \theta} \ln \left\{ \det \left[ H\left(\frac{\theta}{L}, \epsilon\right) - E_B \right] \right\}.
\end{equation}
This topological invariant quantifies the total phase $\Phi $, defined as the argument of $\det[H(\theta/L,\epsilon)-E_B]$, accumulated by the complex eigenvalues as the flux $\theta$ evolves over a $\kappa$-period scaled by the system size \cite{13}. Due to the $\kappa$-periodic modulation of the mosaic potential, the fundamental periodicity of the spectral flow is altered. Consequently, the prefactor $\kappa$ is strictly required to ensure the integer quantization of $\omega$.

The physical interpretation of $\omega$ is rooted in the duality between real and momentum space. Eigenstates that are localized in real space correspond to extended states in the dual (momentum) space. These extended states are highly sensitive to the phase $\theta$, causing their eigenvalues to traverse closed loops in the complex plane, thereby generating a non-zero winding number ($\omega = 1$) if $E_B$ lies within the loop. Conversely, extended eigenstates in real space are localized in the dual space and remain largely insensitive to the boundary condition. Their eigenvalues do not exhibit significant spectral flow, resulting in a trivial winding number ($\omega = 0$).

We first present numerical results for $\kappa=2$ and $\epsilon=0.8$ to illustrate the evolution of the energy spectrum. As shown in Fig.~\ref{fig: placeholder2}, the complex energy spectrum undergoes a dramatic transformation as the potential strength $V$ increases. In the fully extended phase ($V < V_1$), the spectrum exhibits a cross-shaped structure. In the intermediate ME phase ($V_1 < V$), localized and extended states coexist, and the spectrum bifurcates into distinct regions.

The topological signature of these phases is captured by the winding number. To probe distinct regions of the complex plane, we select a base energy $E_{B} = 1.27 + i(V-0.6)$ and compute the corresponding winding number $\omega$. As illustrated in Fig.~\ref{fig:placeholder3}(a), the evolution of $\omega$ with respect to $V$ serves as a robust order parameter. This behavior clearly delineates the transition of the base point $E_B$ from the extended phase to the localized phase as $V$ increases.

The geometric origin of this quantization is elucidated in Figs.~\ref{fig:placeholder3} (b). For a base energy in a localized region (e.g., $E_{B} = 1.27 + 1.5i$ at $V=2.1$), the accumulated phase of the determinant $\Phi $ exhibits a linear dependence on $\theta$ with a slope of unity, confirming a net phase winding of $\pi$. This behavior is in full agreement with the theoretical prediction that point-gap topology protects the localized boundary modes in NH quasiperiodic systems.

\section{Conclusions}
\label{Conclusions}

In summary, we have proposed and analyzed the non-Hermitian mosaic Maryland model, a system that integrates the singular nature of the Maryland potential with discrete structural modulation. By leveraging Avila's global theory, we overcame the challenges posed by the loss of integrability to derive an exact analytical expression for $\Gamma(E)$. This theoretical framework enabled us to formulate a precise analytical condition for the mobility edges in the complex energy plane for arbitrary mosaic spacing $\kappa$.

Focusing on the representative case of $\kappa=2$, we mapped out a rich phase diagram characterized by two distinct regimes: a fully localized phase, an intermediate ME phase containing coexisting localized and extended states. We established the spectral winding number under periodic boundary conditions as a robust topological order parameter, which successfully demarcates these dynamical phases. The analytical predictions were rigorously corroborated by comprehensive numerical calculations---including the IPR, NPR, and FD---which confirmed the critical threshold $V_1$ and the nature of the eigenstates in each regime. Our findings highlight the potential of discrete structural modulation as a powerful degree of freedom for engineering wave confinement and transport in NH systems.  We hope this work stimulates further theoretical investigations into these critical boundaries and inspires experimental realizations in synthetic photonic or acoustic lattices.

\section{ACKNOWLEDGMENTS}
This work was supported by Natural Science Foundation of Jiangsu Province (Grant No. BK20231320).

\renewcommand{\appendixname}{APPENDIX}
\appendix
\section{Lyapunov Exponent Analysis}
\label{Appendix A}
The spectral characteristics of the system are governed by the eigenvalue equation:
\begin{equation}\label{eq:eigenvalue_problem}
    \psi_{n+1} + \psi_{n-1} + V_n \psi_n = E\psi_n.
\end{equation}
The core feature of this model is the NH quasiperiodic mosaic potential, defined as:
\begin{equation}\label{eq:mosaic_potential}
V_n =
\begin{cases}
V \tan(\pi\alpha n + \theta + i\epsilon), & n = m\kappa, \\
0, & \text{otherwise},
\end{cases}
\end{equation}
where $m \in \mathbb{Z}$. To analytically determine the mobility edges, we compute the $\Gamma(E)$ using Avila's global theory for analytic $SL(2, \mathbb{C})$ cocycles \cite{97,82,47}. The LE is defined by the asymptotic growth rate of the transfer matrix norm:
\begin{equation}\label{eq:LE_def}
    \Gamma(E) = \lim_{N \to \infty} \frac{1}{N\kappa} \ln \left\| \prod_{m=1}^{N} T_m \right\|,
\end{equation}
where $T_m$ is the monodromy matrix associated with the $m$-th unit cell of size $\kappa$. This matrix is constructed from the product of one interaction matrix (at the potential site) and $\kappa-1$ free-propagation matrices:
\begin{equation}\label{eq:Tm_def}
    T_m =
    \begin{pmatrix}
    E - V_m & -1 \\
    1 & 0
    \end{pmatrix}
    \begin{pmatrix}
    E & -1 \\
    1 & 0
    \end{pmatrix}^{\kappa-1},
\end{equation}
where $\mathcal{V}_m \equiv V \tan(\pi \alpha m \kappa + \theta + i\epsilon)$.

To facilitate the application of Avila's global theory, we regularize the singularity of the tangent potential by decomposing $T_m$ as $T_m = Y_m / X_m$. Here, we define the scalar factor $X_m = \cos(\pi \alpha m \kappa + \theta + i\epsilon)$ and the regularized matrix $Y_m$ as:
\begin{equation}\label{eq:Ym_def}
Y_m =
\begin{pmatrix}
E X_m - \tilde{\mathcal{V}}_m & -X_m \\
X_m & 0
\end{pmatrix}
\begin{pmatrix}
a_\kappa & -a_{\kappa-1} \\
a_{\kappa-1} & -a_{\kappa-2}
\end{pmatrix},
\end{equation}
where $\tilde{V}_m = V \sin(\pi \alpha m \kappa + \theta + i\epsilon)$. In the above derivation, we have utilized the identity for the $(\kappa-1)$-th power of the transfer matrix:
\begin{equation}\label{eq:transfer_matrix_power}
\begin{pmatrix}
E & -1 \\
1 & 0
\end{pmatrix}^{\kappa-1} =
\begin{pmatrix}
a_\kappa & -a_{\kappa-1} \\
a_{\kappa-1} & -a_{\kappa-2}
\end{pmatrix},
\end{equation}
where $a_\kappa$ represents the modified Chebyshev polynomials of the second kind, given by:
\begin{equation}\label{eq:ak_def}
    a_\kappa = \frac{1}{D} \left[ \left(\frac{E+D}{2}\right)^\kappa - \left(\frac{E-D}{2}\right)^\kappa \right],
\end{equation}
with $D = \sqrt{E^2 - 4}$.

\begin{figure*}[t]
    \centering
    \includegraphics[width=1\linewidth]{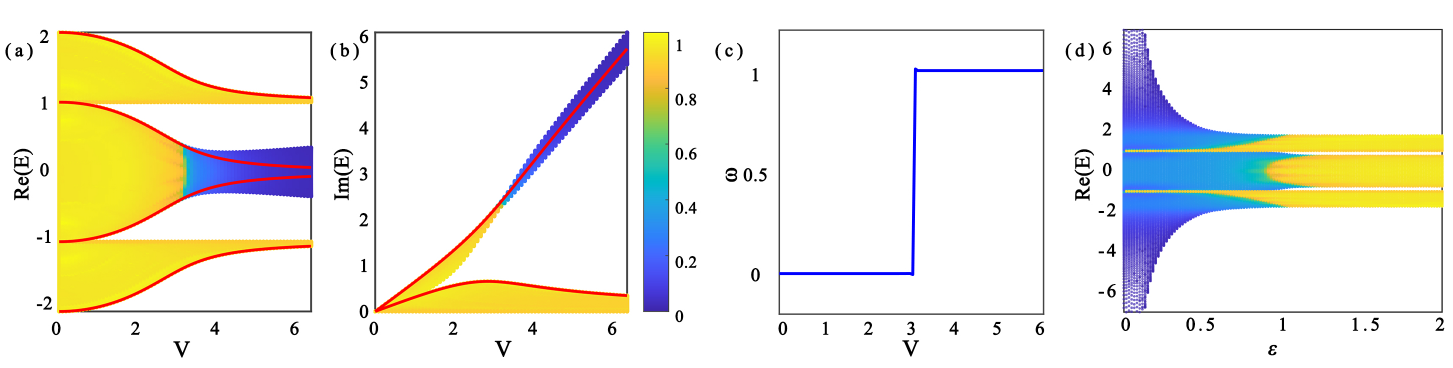}
    \caption{Comparison of analytical mobility edges and numerical localization landscapes for the mosaic modulation $\kappa = 3$. 
    (a) Real part of the energy spectrum $\text{Re}(E)$ as a function of the potential strength $V$. The color scale represents the (FD). The red solid curves denote the analytical mobility edges derived from Eq.~(\ref{eq:ME_kappa3}). Note the two robust extended bands manifesting as horizontal lines at $\text{Re}(E) = \pm 1$. 
    (b) Imaginary part of the energy spectrum $\text{Im}(E)$ versus $V$. The analytical red curves accurately bound the numerically obtained spectrum for a relatively large $\epsilon$.
    (c) Topological winding number $\omega$ as a function of $V$, computed with a reference base energy at $\text{Re}(E) = 0$. The abrupt transition from $\omega=0$ to $\omega=1$ at $V = 3.07$. System parameters are set as $L=987$ and $\epsilon=1.8$.
    (d) Evolution of the real part of the energy spectrum $Re(E)$ as a function of the non-Hermitian parameter $\epsilon$. The horizontal persistent bands at $Re(E)=\pm1$ maintain a high FD (yellow) across all plotted values of $\epsilon$.
    }
    \label{fig:placeholder4}
\end{figure*}

Connection to Numerical Spectra: The polynomials $a_\kappa(E)$ govern both the analytical Lyapunov exponent and the numerical energy spectra. For instance, the condition $a_\kappa(E) = 0$ often signals the presence of extended states, as observed in the persistent extended bands at specific energies in Fig.~\ref{fig:placeholder1}(a) for $\kappa=2$ and later in Fig.~\ref{fig:placeholder4}(a) for $\kappa=3$.

Substituting the decomposition into Eq.~(\ref{eq:LE_def}), the LE can be expressed as:
\begin{equation}\label{eq:LE_split}
    \Gamma(E) = \lim_{N \to \infty} \frac{1}{N\kappa} \left[ \ln \left\| \prod_{m=1}^{N} Y_m \right\| - \sum_{m=1}^{N} \ln |X_m| \right].
\end{equation}
The second term in Eq.~(\ref{eq:LE_split}) is evaluated by integrating over the ergodic phase variable $\varphi = \pi \alpha n + \theta$:
\begin{equation}
\lim_{N \to \infty} \frac{1}{N\kappa} \sum_{m=1}^{N} \ln |X_m| = \frac{1}{\pi \kappa} \int_{-\pi/2}^{\pi/2} \ln |\cos \varphi| \, d\varphi = -\frac{\ln 2}{\kappa}.
\end{equation}

Avila's global theory establishes that $\Gamma(\epsilon)$ is a convex, piecewise linear function of $\epsilon$. Therefore, the behavior of $\Gamma(E)$ for generic $\epsilon$ is determined by its asymptotic form in the large non-Hermitian limit. We consider the analytic extension $\theta \to \theta + i\epsilon$ and take the limit $\epsilon \to \infty$. In this regime, the trigonometric terms behave as:
\begin{equation}
X_m \approx \frac{e^{\epsilon} e^{-i(\pi \alpha m \kappa + \theta)}}{2}, \quad \tilde{\mathcal{V}}_m \approx \frac{V e^{\epsilon} e^{-i(\pi \alpha m \kappa + \theta)}}{2i}.
\end{equation}
Consequently, the regularized matrix $Y_m$ simplifies to:
\begin{equation}
Y_m(\epsilon) \approx \frac{e^{\epsilon} e^{-i(\pi \alpha m \kappa + \theta)}}{2i}
\begin{pmatrix}
E - iV & -1 \\
1 & 0
\end{pmatrix}
\begin{pmatrix}
a_\kappa & -a_{\kappa-1} \\
a_{\kappa-1} & -a_{\kappa-2}
\end{pmatrix}.
\end{equation}
Defining the complex parameter $\chi \equiv E - iV$, we obtain the compact form:
\begin{equation}
Y_m(\epsilon) \approx \frac{e^{\epsilon} e^{-i(\pi \alpha m \kappa + \theta)}}{2i}
\begin{pmatrix}
\chi a_\kappa - a_{\kappa-1} & -\chi a_{\kappa-1} + a_{\kappa-2} \\
a_\kappa & -a_{\kappa-2}
\end{pmatrix}.
\end{equation}
Finally, substituting this asymptotic form back into the LE expression yields:
\begin{equation}
\lim_{N \to \infty} \frac{1}{N\kappa} \ln \left\| \prod_{m=1}^{N} Y_m \right\| = \frac{\epsilon}{\kappa} + \frac{\ln f}{\kappa},
\end{equation}
where the spectral factor $f$ is given by the maximum eigenvalue of the core matrix:
\begin{equation}\label{eq:f_def}
f = \max_{\pm} \left| \frac{\chi a_\kappa - 2a_{\kappa-1} \pm G}{4} \right|,
\end{equation}
with the discriminant
\begin{equation}\label{eq:G_def}
G = \sqrt{\chi^2 a_\kappa^2 - 4\chi a_\kappa a_{\kappa-1} + 4a_\kappa a_{\kappa-2}}.
\end{equation}
Using the asymptotic result derived above and the expression for the second term, we obtain the limiting behavior $\Gamma(E) \approx \frac{1}{\kappa}(\epsilon + \ln 2f)$. According to Avila's global theory, the Lyapunov exponent $\kappa \Gamma(E, \epsilon)$ is a convex, piecewise linear function of $\epsilon$ with quantized integer slopes. This implies that the function must take the form $\kappa \Gamma(E) = \max \{ \epsilon + \ln 2f, \kappa \Gamma_0(E) \}$ \cite{44}. Crucially, for energies $E$ belonging to the continuous spectrum, the Lyapunov exponent must vanish. Since $\Gamma_\epsilon(E)$ behaves as an affine function in the vicinity of $\epsilon=0$, and enforcing the non-negativity condition $\Gamma(E) \ge 0$, we arrive at the exact analytical expression valid for all $E$ in the spectrum:
\begin{equation}\label{eq:Gamma_final}
\Gamma(E) = \frac{1}{\kappa} \max \{ \ln(2f), 0 \}.
\end{equation}

We now apply this general result to the specific case of $\kappa=2$. The spectral polynomials reduce to $a_0=0$, $a_1=1$, and $a_2=E$, which simplifies the discriminant $G$ to:
\begin{equation}
G = \sqrt{\chi^2 E^2 - 4\chi E} = \sqrt{(\chi E - 2)^2 - 4}.
\end{equation}
The mobility edge is precisely defined by the vanishing condition $\Gamma(E)=0$, which is equivalent to $2f=1$. Substituting the expression for $f$, this condition leads to:
\begin{equation}
\max_{\pm} \left| (\chi E - 2) \pm \sqrt{(\chi E - 2)^2 - 4} \right| = 2.
\end{equation}
Utilizing the algebraic identity for inverse hyperbolic functions, this equation simplifies to the compact analytical form:
\begin{equation}\label{eq:ME_analytic}
\left| \chi E - 2 \right| = 2,
\end{equation}
where $\chi = E - iV$. The solutions to this equation describe the boundary between localized and extended states in the complex plane, which is vividly captured in the numerical spectra of Fig.~\ref{fig:placeholder1}(a). At the same time,this formula can be used to determine the imaginary part of the energy spectrum as a function of $V$ which can also be captured in the spectra of Fig.\ref{fig:placeholder1}(b).

\section{Analytical Mobility Edges for Higher-Order Modulations ($\kappa=3, 4$)}
\label{Appendix B}

Building upon the general expression for the Lyapunov exponent derived in Eq.~(\ref{eq:Gamma_final}), we now explicitly determine the analytical forms of the mobility edges for higher-order mosaic modulations, specifically $\kappa=3$ and $\kappa=4$. The condition for the mobility edge, $\Gamma(E)=0$ (or equivalently $2f=1$), simplifies in the asymptotic limit to the algebraic constraint:
\begin{equation}
    \left| \chi a_\kappa - 2a_{\kappa-1} \right| = 2.
\end{equation}
We apply this criterion to each case below.

\subsubsection{Case $\kappa = 3$}
For a modulation period of $\kappa=3$, the relevant spectral polynomials are obtained from the recurrence relations as $a_1=1$, $a_2=E$, and $a_3 = E^2 - 1$. Substituting these into the general criterion yields the discriminant form:
\begin{equation}
    G_3 = \sqrt{\left[ \chi(E^2-1) - 2E \right]^2 - 4}.
\end{equation}
Consequently, the exact analytical ME is given by:
\begin{equation}\label{eq:ME_kappa3}
    \left| \chi (E^2 - 1) - 2E \right| = 2.
\end{equation}
The analytical boundary perfectly captures the evolution of imaginary part of eigenenergy as a function of $V$, which is numerically corroborated in Figs.~\ref{fig:placeholder4}(a) and (b). To topologically characterize this phase transition, we compute the winding number $\omega$ for a base energy at $\text{Re}(E) = 0$. As shown in Fig.~\ref{fig:placeholder4}(c), $\omega$ exhibits an abrupt jump from $0$ to $1$ at $V \approx 2.7$, perfectly marking the onset of the localized phase in excellent agreement with the analytical ME.

\begin{figure*}[t]
    \centering
    \includegraphics[width=1\linewidth]{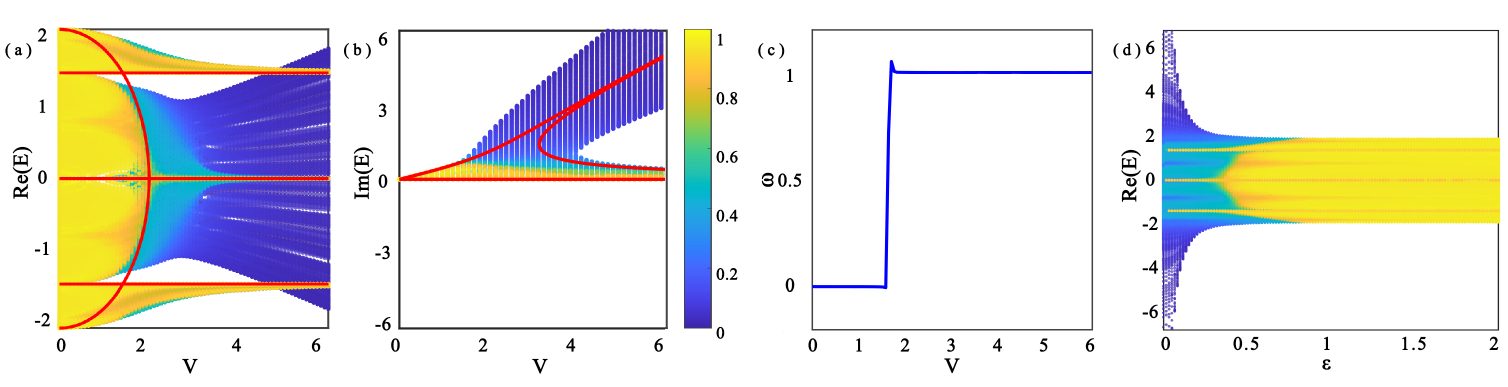}
    \caption{Comparison of analytical mobility edges and numerical localization landscapes for the mosaic modulation $\kappa = 4$. 
    (a) Phase diagram in the $V-\text{Re}(E)$ plane, with the color scale encoding the FD. The red solid lines map the exact analytical mobility edges obtained from Eq.~(\ref{eq:ME_kappa4}). Three persistent extended bands are clearly visible at $\text{Re}(E) = 0$ and $\pm\sqrt{2}$. 
    (b) Phase diagram in the $V-\text{Im}(E)$ plane. 
    (c) Winding number $\omega$ versus $V$, computed with a reference base energy at $\text{Re}(E) = 0.3$. The topological phase transition occurs at $V = 1.74$. System parameters are set as $L=1000$ and $\epsilon=0.8$.
    (d) Energy spectrum $Re(E)$ versus the non-Hermitian parameter $\epsilon$. The three extended bands at $Re(E)=0$ and $\pm\sqrt{2}$ remain perfectly stable and delocalized, independent of the non-Hermitian effect.
    }
    \label{fig:placeholder5}
\end{figure*}
A distinct feature of this solution arises at the roots of $a_3(E)=0$, specifically $E = \pm 1$. At these energy values, the condition for localization breaks down structurally, resulting in two robust extended bands that persist regardless of the potential strength $V$. As depicted in the $V-\text{Re}(E)$ phase diagram [see Fig.~\ref{fig:placeholder4}(a)], these robust bands manifest as two persistent horizontal straight lines at $\text{Re}(E) = \pm 1$.
Furthermore, to explicitly verify their independence from the non-Hermiticity, we plot the energy spectrum as a function of $\epsilon$ in Fig.~\ref{fig:placeholder4}(d)]. Strikingly, while the bulk of the spectrum undergoes severe deformations and localization transitions as $\epsilon$ increases, the resonant states at $Re(E)=\pm1$ retain a fractal dimension near unity (yellow region), proving their absolute robustness against non-Hermitian dissipation.

\subsubsection{Case $\kappa = 4$}
Similarly, for $\kappa=4$, the spectral polynomials are $a_3 = E^2-1$ and $a_4 = E(E^2-2) = E^3 - 2E$. Inserting these explicitly into the general formula leads to:
\begin{equation}
    G_4 = \sqrt{\left[ \chi(E^3-2E) - 2(E^2-1) \right]^2 - 4}.
\end{equation}
The corresponding ME equation is given by:
\begin{equation}\label{eq:ME_kappa4}
    \left| \chi (E^3 - 2E) - 2(E^2 - 1) \right| = 2.
\end{equation}
Similarly, the exact ME formula successfully predicts the complex spectral boundaries in both the $V-\text{Re}(E)$ and $V-\text{Im}(E)$ planes, as shown in Figs.~\ref{fig:placeholder5} (a) and (b). Consistent with these spectral features, the topological transition is perfectly captured by $\omega$ [see Fig.~\ref{fig:placeholder5}(c)]. Evaluated at a reference energy with $\text{Re}(E) = 0.1$, the winding number sharply transitions from $0$ to $1$ at $V \approx 1.85$, structurally corroborating the critical boundary derived from Eq.~(\ref{eq:ME_kappa4}).

The system supports three robust extended bands located at the roots of $a_4(E)=0$, namely $E=0$ and $E=\pm\sqrt{2}$. In the computed phase diagram [see Fig.~\ref{fig:placeholder5}(a)], these manifest as three distinct horizontal lines with FD approaching $1$, confirming that the number of such robust bands strictly follows the $\kappa-1$ rule.
This extraordinary structural protection is further corroborated in Fig.~\ref{fig:placeholder5}(d), where the delocalized nature of these three specific bands ($Re(E)=0, \pm\sqrt{2}$) remains completely unperturbed even in the large $\epsilon$ limit, sharply contrasting with the rest of the vulnerable energy spectrum.

\bibliography{reference}

@article{1,
  title = {Non-Hermitian Maryland model},
  author = {Longhi, Stefano},
  journal = {Phys. Rev. B},
  volume = {103},
  issue = {22},
  pages = {224206},
  numpages = {13},
  month = {Jun},
  publisher = {American Physical Society},
  year = {2021},
  url ={https://link.aps.org/doi/10.1103/PhysRevB.103.224206},
  }

@article{2,
  title = {Anderson-skin dualism: A boundary-dependent effect in non-Hermitian disordered coupled systems},
  author = {Li, Shan-Zhong and Li, Linhu and Zhu, Shi-Liang and Li, Zhi},
  journal = {Phys. Rev. B},
  volume = {112},
  issue = {20},
  pages = {L201108},
  numpages = {6},
  month = {Nov},
  publisher = {American Physical Society},
  year = {2025},
  url = {https://link.aps.org/doi/10.1103/vz12-2qn3}

}

@article{4,
  title = {Critical dynamics and its interferometry in the one-dimensional $p$-wave-paired Aubry-Andr\'e-Harper model},
  author = {Zhang, Zhi-Han and Kou, Han-Chuan and Li, Peng},
  journal = {Phys. Rev. B},
  volume = {112},
  issue = {1},
  pages = {014310},
  numpages = {11},
  month = {Jul},
  publisher = {American Physical Society},
  year = {2025},
  url = {https://link.aps.org/doi/10.1103/6331-k7vl}
 
}

@article{5,
  title = {Fundamental localized quantum state in disordered systems: The balance state},
  author = {Bai, Xiao-Dong and Fu, Jinyuan and Li, Tiantian and Wang, Ji-Guo and Zhao, Jincui},
  journal = {Phys. Rev. B},
  volume = {111},
  issue = {22},
  pages = {224206},
  numpages = {7},
  month = {Jun},
  publisher = {American Physical Society},
  year = {2025},
  url = {https://link.aps.org/doi/10.1103/8qhf-yvsk}
}

@article{7,
      title={The fundamental localization phases in quasiperiodic systems: A unified framework and exact results}, 
      author={Xin-Chi Zhou and Bing-Chen Yao and Yongjian Wang and Yucheng Wang and Yudong Wei and Qi Zhou and Xiong-Jun Liu},
      
     journal={arXiv},
     volume={2503},
     pages={24380},
      year={2025},
        url={https://arxiv.org/abs/2503.24380}, 
}

@article{8,

url = {https://doi.org/10.1088/1555-6611/adf833},
month = {Aug},
publisher = {IOP Publishing},
volume = {35},
number = {8},
pages = {086202},
author = {Wang, Dijie and Zhang, Ruohan and Wang, Zhengling and Wahab, Abdul},
title = {Achieving optical signal switching based on non-Hermitian topological states},
journal = {Laser Phys.},
year = {2025}
}

@Article{9,
author = {Ding, Wenchen and Feng, Yaru},
title = {Second-Order Topological States in Non-Hermitian Square Photonic Crystals},
journal = {Photonics},
volume = {12},
number = {11},
ARTICLE-NUMBER = {1087},
year = {2025},
ISSN = {2304-6732},
URL = {https://www.mdpi.com/2304-6732/12/11/1087},
}

@article{24,
  title = {One-Dimensional Quasiperiodic Mosaic Lattice with Exact Mobility Edges},
  author = {Wang, Yucheng and Xia, Xu and Zhang, Long and Yao, Hepeng and Chen, Shu and You, Jiangong and Zhou, Qi and Liu, Xiong-Jun},
  journal = {Phys. Rev. Lett.},
  volume = {125},
  issue = {19},
  pages = {196604},
  numpages = {6},
  month = {Nov},
  publisher = {American Physical Society},
  year = {2020},
   url = {https://link.aps.org/doi/10.1103/PhysRevLett.125.196604}
}

@article{11,
  title = {Topological Phases of Non-Hermitian Systems},
  author = {Gong, Zongping and Ashida, Yuto and Kawabata, Kohei and Takasan, Kazuaki and Higashikawa, Sho and Ueda, Masahito},
  journal = {Phys. Rev. X},
  volume = {8},
  issue = {3},
  pages = {031079},
  numpages = {33},
  month = {Sep},
  publisher = {American Physical Society},
  year = {2018},
  url = {https://link.aps.org/doi/10.1103/PhysRevX.8.031079}
}

@article{13,
  title = {Non-Hermitian Topological Invariants in Real Space},
  author = {Song, Fei and Yao, Shunyu and Wang, Zhong},
  journal = {Phys. Rev. Lett.},
  volume = {123},
  issue = {24},
  pages = {246801},
  numpages = {8},
  month = {Dec},
  publisher = {American Physical Society},
  year = {2019},
   url = {https://link.aps.org/doi/10.1103/PhysRevLett.123.246801}
}

@article{14,
month = {Jun},
publisher = {IOP Publishing},
volume = {40},
number = {4},
pages = {045403},
author = {Domínguez-Castro, G A and Paredes, R},
title = {The Aubry–André model as a hobbyhorse for understanding the localization phenomenon},
journal = {Eur. J. Phys.},
year = {2019},
url = {https://doi.org/10.1088/1361-6404/ab1670}
}

@article{17,
  title = {Absence of Diffusion in Certain Random Lattices},
  author = {Anderson, P. W.},
  journal = {Phys. Rev.},
  volume = {109},
  issue = {5},
  pages = {1492--1505},
  numpages = {0},
  month = {Mar},
  publisher = {American Physical Society},
  year = {1958},
   url = {https://link.aps.org/doi/10.1103/PhysRev.109.1492}
}

@article{21,
  title = {Mobility Edge in a Model One-Dimensional Potential},
  author = {Das Sarma, S. and He, Song and Xie, X. C.},
  journal = {Phys. Rev. Lett.},
  volume = {61},
  issue = {18},
  pages = {2144--2147},
  numpages = {0},
  year = {1988},
  month = {Oct},
  publisher = {American Physical Society},
  doi = {10.1103/PhysRevLett.61.2144},
  url = {https://link.aps.org/doi/10.1103/PhysRevLett.61.2144}
}

@article{22,
  title = {Localization in one-dimensional incommensurate lattices beyond the Aubry-Andr\'e model},
  author = {Biddle, J. and Wang, B. and Priour, D. J. and Das Sarma, S.},
  journal = {Phys. Rev. A},
  volume = {80},
  issue = {2},
  pages = {021603},
  numpages = {4},
  year = {2009},
  month = {Aug},
  publisher = {American Physical Society},
  doi = {10.1103/PhysRevA.80.021603},
  url = {https://link.aps.org/doi/10.1103/PhysRevA.80.021603}
}

@article{23,
  title = {Predicted Mobility Edges in One-Dimensional Incommensurate Optical Lattices: An Exactly Solvable Model of Anderson Localization},
  author = {Biddle, J. and Das Sarma, S.},
  journal = {Phys. Rev. Lett.},
  volume = {104},
  issue = {7},
  pages = {070601},
  numpages = {4},
  year = {2010},
  month = {Feb},
  publisher = {American Physical Society},
  doi = {10.1103/PhysRevLett.104.070601},
  url = {https://link.aps.org/doi/10.1103/PhysRevLett.104.070601}
}

@article{27,
   title={Exact Mobility Edges for 1D Quasiperiodic Models},
   volume={401},
   ISSN={1432-0916},
   url={http://dx.doi.org/10.1007/s00220-023-04695-9},
   DOI={10.1007/s00220-023-04695-9},
   number={3},
   journal = {Commun. Math. Phys.},
   publisher={Springer Science and Business Media LLC},
   author={Wang, Yongjian and Xia, Xu and You, Jiangong and Zheng, Zuohuan and Zhou, Qi},
   year={2023},
   month={Mar}, 
   pages={2521–2567}, 
   }

@article{28,
      title={Anti-resonances and sharp analysis of Maryland localization for all parameters}, 
      author={Rui Han and Svetlana Jitomirskaya and Fan Yang},
     journal={arXiv},
      volume={2205},
      pages={04021},
      YEAR={2022},
       url={https://arxiv.org/abs/2205.04021}, 
}

@article{30,
 ISSN = {0003486X, 19398980},
 URL = {http://www.jstor.org/stable/121066},
 author = {Svetlana Ya. Jitomirskaya},
 journal = {Ann. Math.},
 number = {3},
 pages = {1159--1175},
 publisher = {[Annals of Mathematics, Trustees of Princeton University on Behalf of the Annals of Mathematics, Mathematics Department, Princeton University]},
 title = {Metal-Insulator Transition for the Almost Mathieu Operator},
 urldate = {2026-01-30},
 volume = {150},
 year = {1999}
}

@article{31,
author = {Jitomirskaya, Svetlana and Liu, Wencai},
title = {Arithmetic Spectral Transitions for the Maryland Model},
journal = {Commun. Pure Appl. Math.},
volume = {70},
number = {6},
pages = {1025-1051},
doi = {https://doi.org/10.1002/cpa.21688},
year = {2017}
}

@article{32, 
title={Pure point spectrum for the Maryland model: a constructive proof}, volume={41}, DOI={10.1017/etds.2019.50}, number={1}, journal = {Ergodic Theory Dyn. Syst.}, author={JITOMIRSKAYA, SVETLANA and YANG, FAN}, year={2021}, pages={283–294}}

@article{34,
doi = {10.1088/1361-6633/ad4e64},
url = {https://doi.org/10.1088/1361-6633/ad4e64},
year = {2024},
month = {Jul},
publisher = {IOP Publishing},
volume = {87},
number = {7},
pages = {078002},
author = {Yang, Kang and Li, Zhi and König, J Lukas K and Rødland, Lukas and Stålhammar, Marcus and Bergholtz, Emil J},
title = {Homotopy, symmetry, and non-Hermitian band topology},
journal = {Rep. Prog. Phys.}
}

@article{39,
    author = {He, Jiawei and Xia, Xu},
    title = {Arithmetic phase transitions for mosaic Maryland model},
    journal = {J. Math. Phys.},
    volume = {64},
    number = {4},
    pages = {043504},
    year = {2023},
    month = {Apr},
    issn = {0022-2488},
    doi = {10.1063/5.0123576},
   
}

@article{40,
doi = {10.1088/1361-648X/ac4530},
year = {2022},
month = {Jan},
publisher = {IOP Publishing},
volume = {34},
number = {11},
pages = {115402},
author = {Zhou, Longwen and Gu, Yongjian},
title = {Topological delocalization transitions and mobility edges in the nonreciprocal Maryland model},
journal = {J. Phys.: Condens. Matter}
}

@article{41,
  title = {Ring structure in the complex plane: A fingerprint of a non-Hermitian mobility edge},
  author = {Li, Shan-Zhong and Li, Zhi},
  journal = {Phys. Rev. B},
  volume = {110},
  issue = {4},
  pages = {L041102},
  numpages = {8},
  year = {2024},
  month = {Jul},
  publisher = {American Physical Society},
  doi = {10.1103/PhysRevB.110.L041102},
  
}

@article{42,
author = {Jitomirskaya, Svetlana and Liu, Wencai},
title = {Arithmetic Spectral Transitions for the Maryland Model},
journal = {Commun. Pure Appl. Math.},
volume = {70},
number = {6},
pages = {1025-1051},
doi = {https://doi.org/10.1002/cpa.21688},
year = {2017}
}

@article{43,
  url={http://dx.doi.org/10.1103/PhysRevA.97.052115},
   title={Geometrical meaning of winding number and its characterization of topological phases in one-dimensional chiral non-Hermitian systems},
   volume={97},
   ISSN={2469-9934},
   number={5},
   journal={Phys. Rev. A},
   publisher={American Physical Society (APS)},
   author={Yin, Chuanhao and Jiang, Hui and Li, Linhu and Lü, Rong and Chen, Shu},
   year={2018},
   month=may }

@article{44,
  title = {Exact mobility edges, $\mathcal{PT}$-symmetry breaking, and skin effect in one-dimensional non-Hermitian quasicrystals},
  author = {Liu, Yanxia and Wang, Yucheng and Liu, Xiong-Jun and Zhou, Qi and Chen, Shu},
  journal = {Phys. Rev. B},
  volume = {103},
  issue = {1},
  pages = {014203},
  numpages = {9},
  year = {2021},
  month = {Jan},
  publisher = {American Physical Society},
  doi = {10.1103/PhysRevB.103.014203},
  url = {https://link.aps.org/doi/10.1103/PhysRevB.103.014203}
}

@article{45,
  title = {Hidden Self Duality and Exact Mobility Edges in Quasiperiodic Network Models},
  author = {Hu, Hai-Tao and Lin, Xiaoshui and Guo, Ai-Min and Guo, Guangcan and Lin, Zijing and Gong, Ming},
  journal = {Phys. Rev. Lett.},
  volume = {134},
  issue = {24},
  pages = {246301},
  numpages = {8},
  year = {2025},
  month = {Jun},
  publisher = {American Physical Society},
  doi = {10.1103/rl1f-ptzq},
  url = {https://link.aps.org/doi/10.1103/rl1f-ptzq}
}

@article{47,
  title = {Asymmetric transfer matrix analysis of Lyapunov exponents in one-dimensional nonreciprocal quasicrystals},
  author = {Li, Shan-Zhong and Cheng, Enhong and Zhu, Shi-Liang and Li, Zhi},
  journal = {Phys. Rev. B},
  volume = {110},
  issue = {13},
  pages = {134203},
  numpages = {11},
  year = {2024},
  month = {Oct},
  publisher = {American Physical Society},
  doi = {10.1103/PhysRevB.110.134203},
  url = {https://link.aps.org/doi/10.1103/PhysRevB.110.134203}
}

@article{49,
doi = {10.1088/0034-4885/70/6/R03},

month = {May},
publisher = {},
volume = {70},
number = {6},
pages = {947},
author = {Bender, Carl M},
title = {Making sense of non-Hermitian Hamiltonians},
journal = {Rep. Prog. Phys.},
year = {2007}
}

@article{50,
  title = {Winding numbers and generalized mobility edges in non-Hermitian systems},
  author = {Zeng, Qi-Bo and Xu, Yong},
  journal = {Phys. Rev. Res.},
  volume = {2},
  issue = {3},
  pages = {033052},
  numpages = {7},
  year = {2020},
  month = {Jul},
  publisher = {American Physical Society},
  doi = {10.1103/PhysRevResearch.2.033052},
  url = {https://link.aps.org/doi/10.1103/PhysRevResearch.2.033052}
}

@article{51,
  title = {Topological Phase Transitions and Mobility Edges in Non-Hermitian Quasicrystals},
  author = {Lin, Quan and Li, Tianyu and Xiao, Lei and Wang, Kunkun and Yi, Wei and Xue, Peng},
  journal = {Phys. Rev. Lett.},
  volume = {129},
  issue = {11},
  pages = {113601},
  numpages = {6},
  year = {2022},
  month = {Sep},
  publisher = {American Physical Society},
  doi = {10.1103/PhysRevLett.129.113601},
  url = {https://link.aps.org/doi/10.1103/PhysRevLett.129.113601}
}

@article{52,
  title = {Non-Hermitian mobility edges in one-dimensional quasicrystals with parity-time symmetry},
  author = {Liu, Yanxia and Jiang, Xiang-Ping and Cao, Junpeng and Chen, Shu},
  journal = {Phys. Rev. B},
  volume = {101},
  issue = {17},
  pages = {174205},
  numpages = {7},
  year = {2020},
  month = {May},
  publisher = {American Physical Society},
  doi = {10.1103/PhysRevB.101.174205},
  url = {https://link.aps.org/doi/10.1103/PhysRevB.101.174205}
}

@article{53,
   title={Localization and mobility edges in non-Hermitian continuous quasiperiodic systems},
   volume={27},
   ISSN={1367-2630},
   url={http://dx.doi.org/10.1088/1367-2630/adf56f},
   DOI={10.1088/1367-2630/adf56f},
   number={8},
  journal = {New J. Phys.},
   publisher={IOP Publishing},
   author={Jiang, Xiang-Ping and Liu, Zhende and Hu, Yayun and Hou, Hongsheng and Pan, Lei},
   year={2025},
   month={Aug}, 
   pages={083201}, 
   }

@article{54,
  title = {Topological Origin of Non-Hermitian Skin Effects},
  author = {Okuma, Nobuyuki and Kawabata, Kohei and Shiozaki, Ken and Sato, Masatoshi},
  journal = {Phys. Rev. Lett.},
  volume = {124},
  issue = {8},
  pages = {086801},
  numpages = {7},
  year = {2020},
  month = {Feb},
  publisher = {American Physical Society},
  doi = {10.1103/PhysRevLett.124.086801},
  url = {https://link.aps.org/doi/10.1103/PhysRevLett.124.086801}
}

@article{55,
  title = {Correspondence between Winding Numbers and Skin Modes in Non-Hermitian Systems},
  author = {Zhang, Kai and Yang, Zhesen and Fang, Chen},
  journal = {Phys. Rev. Lett.},
  volume = {125},
  issue = {12},
  pages = {126402},
  numpages = {6},
  year = {2020},
  month = {Sep},
  publisher = {American Physical Society},
  doi = {10.1103/PhysRevLett.125.126402},
  url = {https://link.aps.org/doi/10.1103/PhysRevLett.125.126402}
}

@article{57,
  title = {Mobility edge and intermediate phase in one-dimensional incommensurate lattice potentials},
  author = {Li, Xiao and Das Sarma, S.},
  journal = {Phys. Rev. B},
  volume = {101},
  issue = {6},
  pages = {064203},
  numpages = {16},
  year = {2020},
  month = {Feb},
  publisher = {American Physical Society},
  doi = {10.1103/PhysRevB.101.064203},
  url = {https://link.aps.org/doi/10.1103/PhysRevB.101.064203}
}

@article{58,
      title={Exact mobility edges in Aubry-Andr\'{e}-Harper models with relative phases}, 
      author={Xiaoming Cai and Yi-Cong Yu},
     journal={arXiv},
      volume={2205},
      pages={09486},
      year={2022},
      url={https://arxiv.org/abs/2205.09486}, 
}

@article{59,
   title={Non-Hermitian control of localization in mosaic photonic lattices},
   volume={123},
   ISSN={1077-3118},
   url={http://dx.doi.org/10.1063/5.0175675},
   
   number={16},
   journal={Appl. Phys. Lett.},
   publisher={AIP Publishing},
   author={Longhi, Stefano},
   year={2023},
   month={Oct}
   }

@article{60,
  title = {Engineering mobility in quasiperiodic lattices with exact mobility edges},
  author = {Wang, Zhenbo and Zhang, Yu and Wang, Li and Chen, Shu},
  journal = {Phys. Rev. B},
  volume = {108},
  issue = {17},
  pages = {174202},
  numpages = {10},
  year = {2023},
  month = {Nov},
  publisher = {American Physical Society},
  doi = {10.1103/PhysRevB.108.174202},
  url = {https://link.aps.org/doi/10.1103/PhysRevB.108.174202}
}

@article{61,
  title = {Phase transitions in a non-Hermitian Aubry-Andr\'e-Harper model},
  author = {Longhi, Stefano},
  journal = {Phys. Rev. B},
  volume = {103},
  issue = {5},
  pages = {054203},
  numpages = {12},
  year = {2021},
  month = {Feb},
  publisher = {American Physical Society},
  doi = {10.1103/PhysRevB.103.054203},
  url = {https://link.aps.org/doi/10.1103/PhysRevB.103.054203}
}

@article{62,
  title = {Boundary-dependent self-dualities, winding numbers, and asymmetrical localization in non-Hermitian aperiodic one-dimensional models},
  author = {Cai, Xiaoming},
  journal = {Phys. Rev. B},
  volume = {103},
  issue = {1},
  pages = {014201},
  numpages = {12},
  year = {2021},
  month = {Jan},
  publisher = {American Physical Society},
  doi = {10.1103/PhysRevB.103.014201},
  url = {https://link.aps.org/doi/10.1103/PhysRevB.103.014201}
}

@article{64,
  title = {Odd-even effect of the mosaic modulation period of quasiperiodic hopping on the Anderson localization in a one-dimensional lattice model},
  author = {Zhang, Yi-Cai and Yuan, Rong and Song, Shuwei and Hu, Mingpeng and Liu, Chaofei and Wang, Yongjian},
  journal = {Phys. Rev. B},
  volume = {111},
  issue = {6},
  pages = {064201},
  numpages = {10},
  year = {2025},
  month = {Feb},
  publisher = {American Physical Society},
  doi = {10.1103/PhysRevB.111.064201},
  url = {https://link.aps.org/doi/10.1103/PhysRevB.111.064201}
}

@article{63,
  title = {Self-consistent theory of mobility edges in quasiperiodic chains},
  author = {Duthie, Alexander and Roy, Sthitadhi and Logan, David E.},
  journal = {Phys. Rev. B},
  volume = {103},
  issue = {6},
  pages = {L060201},
  numpages = {5},
  year = {2021},
  month = {Feb},
  publisher = {American Physical Society},
  doi = {10.1103/PhysRevB.103.L060201},
  url = {https://link.aps.org/doi/10.1103/PhysRevB.103.L060201}
}

@article{65,
  title = {Emergence of multifractality through cascadelike transitions in a mosaic interpolating Aubry-Andr\'e-Fibonacci chain},
  author = {Dai, Qi and Lu, Zhanpeng and Xu, Zhihao},
  journal = {Phys. Rev. B},
  volume = {108},
  issue = {14},
  pages = {144207},
  numpages = {11},
  year = {2023},
  month = {Oct},
  publisher = {American Physical Society},
  doi = {10.1103/PhysRevB.108.144207},
   url = {https://link.aps.org/doi/10.1103/PhysRevB.108.144207}
}

@article{66,
  title = {Complete delocalization and reentrant topological transition in a non-Hermitian quasiperiodic lattice},
  author = {Padhan, Ashirbad and Padhi, Soumya Ranjan and Mishra, Tapan},
  journal = {Phys. Rev. B},
  volume = {109},
  issue = {2},
  pages = {L020203},
  numpages = {5},
  year = {2024},
  month = {Jan},
  publisher = {American Physical Society},
   url = {https://link.aps.org/doi/10.1103/PhysRevB.109.L020203}
}

@article{67,
  title = {Non-Hermitian robust edge states in one dimension: Anomalous localization and eigenspace condensation at exceptional points},
  author = {Martinez Alvarez, V. M. and Barrios Vargas, J. E. and Foa Torres, L. E. F.},
  journal = {Phys. Rev. B},
  volume = {97},
  issue = {12},
  pages = {121401},
  numpages = {6},
  year = {2018},
  month = {Mar},
  publisher = {American Physical Society},
  doi = {10.1103/PhysRevB.97.121401},
  url = {https://link.aps.org/doi/10.1103/PhysRevB.97.121401}
}

@article{68,
  title = {Skin effect and winding number in disordered non-Hermitian systems},
  author = {Claes, Jahan and Hughes, Taylor L.},
  journal = {Phys. Rev. B},
  volume = {103},
  issue = {14},
  pages = {L140201},
  numpages = {7},
  year = {2021},
  month = {Apr},
  publisher = {American Physical Society},
  doi = {10.1103/PhysRevB.103.L140201},
  url = {https://link.aps.org/doi/10.1103/PhysRevB.103.L140201}
}

@article{69,
  title = {Probing non-Hermitian skin effect and non-Bloch phase transitions},
  author = {Longhi, Stefano},
  journal = {Phys. Rev. Res.},
  volume = {1},
  issue = {2},
  pages = {023013},
  numpages = {13},
  year = {2019},
  month = {Sep},
  publisher = {American Physical Society},
  doi = {10.1103/PhysRevResearch.1.023013},
  url = {https://link.aps.org/doi/10.1103/PhysRevResearch.1.023013}
}

@article{70,
  title = {Anatomy of skin modes and topology in non-Hermitian systems},
  author = {Lee, Ching Hua and Thomale, Ronny},
  journal = {Phys. Rev. B},
  volume = {99},
  issue = {20},
  pages = {201103},
  numpages = {5},
  year = {2019},
  month = {May},
  publisher = {American Physical Society},
  doi = {10.1103/PhysRevB.99.201103},
  url = {https://link.aps.org/doi/10.1103/PhysRevB.99.201103}
}

@article{72,
  title = {Topological phases in non-Hermitian Aubry-Andr\'e-Harper models},
  author = {Zeng, Qi-Bo and Yang, Yan-Bin and Xu, Yong},
  journal = {Phys. Rev. B},
  volume = {101},
  issue = {2},
  pages = {020201},
  numpages = {6},
  year = {2020},
  month = {Jan},
  publisher = {American Physical Society},
  doi = {10.1103/PhysRevB.101.020201},
  url = {https://link.aps.org/doi/10.1103/PhysRevB.101.020201}
}

@article{73,
  title = {Entanglement Phase Transition Induced by the Non-Hermitian Skin Effect},
  author = {Kawabata, Kohei and Numasawa, Tokiro and Ryu, Shinsei},
  journal = {Phys. Rev. X},
  volume = {13},
  issue = {2},
  pages = {021007},
  numpages = {26},
  year = {2023},
  month = {Apr},
  publisher = {American Physical Society},
  doi = {10.1103/PhysRevX.13.021007},
  url = {https://link.aps.org/doi/10.1103/PhysRevX.13.021007}
}

@article{74,
  title = {Unraveling the non-Hermitian skin effect in dissipative systems},
  author = {Longhi, Stefano},
  journal = {Phys. Rev. B},
  volume = {102},
  issue = {20},
  pages = {201103},
  numpages = {6},
  year = {2020},
  month = {Nov},
  publisher = {American Physical Society},
  doi = {10.1103/PhysRevB.102.201103},
  url = {https://link.aps.org/doi/10.1103/PhysRevB.102.201103}
}

@article{75,
  title = {Exceptional topology of non-Hermitian systems},
  author = {Bergholtz, Emil J. and Budich, Jan Carl and Kunst, Flore K.},
  journal = {Rev. Mod. Phys.},
  volume = {93},
  issue = {1},
  pages = {015005},
  numpages = {31},
  year = {2021},
  month = {Feb},
  publisher = {American Physical Society},
  doi = {10.1103/RevModPhys.93.015005},
  url = {https://link.aps.org/doi/10.1103/RevModPhys.93.015005}
}

@article{76,
  title = {Non-Hermitian skin effect and self-acceleration},
  author = {Longhi, Stefano},
  journal = {Phys. Rev. B},
  volume = {105},
  issue = {24},
  pages = {245143},
  numpages = {11},
  year = {2022},
  month = {Jun},
  publisher = {American Physical Society},
  doi = {10.1103/PhysRevB.105.245143},
  url = {https://link.aps.org/doi/10.1103/PhysRevB.105.245143}
}

@article{77,
  title = {Real spectra and phase transition of skin effect in nonreciprocal systems},
  author = {Zeng, Qi-Bo and L\"u, Rong},
  journal = {Phys. Rev. B},
  volume = {105},
  issue = {24},
  pages = {245407},
  numpages = {9},
  year = {2022},
  month = {Jun},
  publisher = {American Physical Society},
  doi = {10.1103/PhysRevB.105.245407},
  url = {https://link.aps.org/doi/10.1103/PhysRevB.105.245407}
}

@article{78,
  title = {Dynamic Signatures of Non-Hermitian Skin Effect and Topology in Ultracold Atoms},
  author = {Liang, Qian and Xie, Dizhou and Dong, Zhaoli and Li, Haowei and Li, Hang and Gadway, Bryce and Yi, Wei and Yan, Bo},
  journal = {Phys. Rev. Lett.},
  volume = {129},
  issue = {7},
  pages = {070401},
  numpages = {6},
  year = {2022},
  month = {Aug},
  publisher = {American Physical Society},
  doi = {10.1103/PhysRevLett.129.070401},
  url = {https://link.aps.org/doi/10.1103/PhysRevLett.129.070401}
}

@article{79,
author = {Yuto Ashida and Zongping Gong and Masahito Ueda},
title = {Non-Hermitian physics},
journal = {Adv. Phys.},
volume = {69},
number = {3},
pages = {249--435},
year = {2020},
publisher = {Taylor \& Francis},
doi = {https://doi.org/10.1080/00018732.2021.1876991},
}

@article{80,
  title = {Classification of topological quantum matter with symmetries},
  author = {Chiu, Ching-Kai and Teo, Jeffrey C. Y. and Schnyder, Andreas P. and Ryu, Shinsei},
  journal = {Rev. Mod. Phys.},
  volume = {88},
  issue = {3},
  pages = {035005},
  numpages = {63},
  year = {2016},
  month = {Aug},
  publisher = {American Physical Society},
  doi = {10.1103/RevModPhys.88.035005},
  url = {https://link.aps.org/doi/10.1103/RevModPhys.88.035005}
}

@article{81,
doi = {10.1088/1361-648X/acd1cb},
url = {https://doi.org/10.1088/1361-648X/acd1cb},
year = {2023},
month = {May},
publisher = {IOP Publishing},
volume = {35},
number = {33},
pages = {333001},
author = {Banerjee, Ayan and Sarkar, Ronika and Dey, Soumi and Narayan, Awadhesh},
title = {Non-Hermitian topological phases: principles and prospects},
journal = {J. Phys.: Condens. Matter}
}

@article{82,
   title = {Non-Hermitian systems and topology: A transfer-matrix perspective},
   volume = {99},
   ISSN = {2469-9969},
   number = {24},
   journal = {Phys. Rev. B},
   publisher = {American Physical Society (APS)},
   author={Kunst, Flore K. and Dwivedi, Vatsal},
   month={Jun},
   year={2019},
   url={http://dx.doi.org/10.1103/PhysRevB.99.245116}
   }

@article{83,
author = {Artur Avila and Jiangong You and Qi Zhou},
title = {Sharp phase transitions for the almost Mathieu operator},
volume = {166},
journal = {Duke Math. J.},
number = {14},
publisher = {Duke University Press},
pages = {2697 -- 2718},
year = {2017},
doi = {10.1215/00127094-2017-0013},
url = {https://doi.org/10.1215/00127094-2017-0013}
}

@article{85,
   title = {Transfer matrix analysis of non-Hermitian Hamiltonians: asymptotic spectra and topological eigenvalues},
   volume={14},
   ISSN = {1664-0403},
   url ={http://dx.doi.org/10.4171/JST/524},
   DOI = {10.4171/jst/524},
   number = {4},
   journal = {J. Spectr. Theory },
   publisher = {European Mathematical Society - EMS - Publishing House GmbH},
   author = {Koekenbier, L. and Schulz-Baldes, H.},
   year = {2024},
   month = {Sep}, 
   pages = {1563–1622} 
   }

@article{87,
    author = {Sanghoon Lee and Tilen Cadez and Kyoung-Min Kim},
    title ={Structural constraints on mobility edges in one-dimensional quasiperiodic systems} ,
    journal = {arXiv},
    volume={2601},
    pages={15799},
    
    year ={2026},
    url={https://arxiv.org/abs/2601.15799}, 

}

@article{88,
    author ={Guolin Nan and Zhijian Li and Feng Mei and Zhihao Xu},
    title ={Anomalous Localization and Mobility Edges in Non-Hermitian Quasicrystals with Disordered Imaginary Gauge Fields} ,
    journal ={arxiv},
    volume={2601},
    pages={14754},
    year = {2026},
    url={https://arxiv.org/abs/2601.14754} 
}

@article{89,
      title={Emergence of distinct exact mobility edges in a quasiperiodic chain}, 
      author={Sanchayan Banerjee and Soumya Ranjan Padhi and Tapan Mishra},
      publisher = {American Physical Society(APS)},
     journal = {Phys. Rev. B},
     
     volume = {111},
      year={2025},
      url = {https://doi.org/10.1103/PhysRevB.111.L220201},
     
}

@article{90,
      title={Coexistence of distinct mobility edges in a 1D quasiperiodic mosaic model}, 
      author={Xu Xia and Weihao Huang and Ke Huang and Xiaolong Deng and Xiao Li},
      journal={arXiv},
      volume={2503},
      pages={04552},
      year={2026},
      url = {https://doi.org/10.48550/arXiv.2503.04552},
      doi = {10.48550/arXiv.2503.04552}
}

@article{92,
  title = {Disordered electronic systems},
  author = {Lee, Patrick A. and Ramakrishnan, T. V.},
  journal = {Rev. Mod. Phys.},
  volume = {57},
  issue = {2},
  pages = {287--337},
  numpages = {0},
  
  month = {Apr},
  year = {1985},
  publisher = {American Physical Society},
  url = {https://link.aps.org/doi/10.1103/RevModPhys.57.287},
  doi = {10.1103/RevModPhys.57.287}
}

@article{93,
doi = {10.1088/0022-3719/20/21/008},
url = {https://doi.org/10.1088/0022-3719/20/21/008},
year = {1987},
month = {Jul},
publisher = {IOP Publishing Ltd},
volume = {20},
number = {21},
pages = {3075},
author = {N Mott},
title = {The mobility edge since 1967},
journal = {J. Phys. C: Solid State Phys.}
}

@article{95,
  title = {Generalized Aubry-Andr\'e self-duality and mobility edges in non-Hermitian quasiperiodic lattices},
  author = {Liu, T. and Guo, H. and Pu, Y. and Longhi, S.},
  journal = {Phys. Rev. B},
  volume = {102},
  issue = {2},
  pages = {024205},
  numpages = {10},
  year = {2020},
  month = {Jul},
  publisher = {American Physical Society},
  doi = {10.1103/PhysRevB.102.024205},
  url = {https://link.aps.org/doi/10.1103/PhysRevB.102.024205}
}

@article{96,
  title = {One-Dimensional Quasiperiodic Mosaic Lattice with Exact Mobility Edges},
  author = {Wang, Y. and Xia, X. and Zhang, L. and Yao, H. and Chen, S. and You, J. and Zhou, Q. and Liu, X.-J.},
  journal = {Phys. Rev. Lett.},
  volume = {125},
  issue = {19},
  pages = {196604},
  numpages = {6},
  year = {2020},
  month = {Nov},
  publisher = {American Physical Society},
  doi = {10.1103/PhysRevLett.125.196604},
  url = {https://link.aps.org/doi/10.1103/PhysRevLett.125.196604}
}

@article{97,
month = {Nov},
publisher = {IOP Publishing},
volume = {35},
number = {3},
pages = {035602},
author = {Cai, Xiaoming and Yu, Yi-Cong},
title = {Exact mobility edges in quasiperiodic systems without self-duality},
journal = {J. Phys.: Condens. Matter},
year = {2022},
doi = {10.1088/1361-648X/aca136}
}

@article{98,
  title = {General approach to the critical phase with coupled quasiperiodic chains},
  author = {Lin, X. and Chen, X. and Guo, G.-C. and Gong, M.},
  journal = {Phys. Rev. B},
  volume = {108},
  issue = {17},
  pages = {174206},
  numpages = {22},
  year = {2023},
  month = {Nov},
  publisher = {American Physical Society},
  doi = {10.1103/PhysRevB.108.174206},
  url = {https://link.aps.org/doi/10.1103/PhysRevB.108.174206}
}

@article{99,
  title = {Fate of localization in a coupled free chain and a disordered chain},
  author = {Lin, X. and Gong, M.},
  journal = {Phys. Rev. A},
  volume = {109},
  issue = {3},
  pages = {033310},
  numpages = {10},
  year = {2024},
  month = {Mar},
  publisher = {American Physical Society},
  doi = {10.1103/PhysRevA.109.033310},
  url = {https://link.aps.org/doi/10.1103/PhysRevA.109.033310}
}

@article{100,
  title = {Anderson transitions},
  author = {Evers, Ferdinand and Mirlin, Alexander D.},
  journal = {Rev. Mod. Phys.},
  volume = {80},
  issue = {4},
  pages = {1355--1417},
  numpages = {0},
  year = {2008},
  month = {Oct},
  publisher = {American Physical Society},
  doi = {10.1103/RevModPhys.80.1355},
  url = {https://link.aps.org/doi/10.1103/RevModPhys.80.1355}
}

@article{101,
  title = {Single-Particle Mobility Edge in a One-Dimensional Quasiperiodic Optical Lattice},
  author = {L\"uschen, Henrik P. and Scherg, S. and Kohlert, T. and Schreiber, M. and Bordia, P. and Li, X. and Das S., S. and Bloch, I.},
  journal = {Phys. Rev. Lett.},
  volume = {120},
  issue = {16},
  pages = {160404},
  numpages = {6},
  year = {2018},
  month = {Apr},
  publisher = {American Physical Society},
  doi = {10.1103/PhysRevLett.120.160404},
  url = {https://link.aps.org/doi/10.1103/PhysRevLett.120.160404}
}

@article{102,
  title = {Metal-Insulator Transition and Scaling for Incommensurate Systems},
  author = {Kohmoto, Mahito},
  journal = {Phys. Rev. Lett.},
  volume = {51},
  issue = {13},
  pages = {1198--1201},
  numpages = {0},
  year = {1983},
  month = {Sep},
  publisher = {American Physical Society},
  doi = {10.1103/PhysRevLett.51.1198},
  url = {https://link.aps.org/doi/10.1103/PhysRevLett.51.1198}
}

@article{103,
  title = {Localization by a Potential with Slowly Varying Period},
  author = {Thouless, D. J.},
  journal = {Phys. Rev. Lett.},
  volume = {61},
  issue = {18},
  pages = {2141--2143},
  numpages = {0},
  year = {1988},
  month = {Oct},
  publisher = {American Physical Society},
  doi = {10.1103/PhysRevLett.61.2141},
  url = {https://link.aps.org/doi/10.1103/PhysRevLett.61.2141}
}

@article{104,
  title = {Interplay of non-Hermitian skin effects and Anderson localization in nonreciprocal quasiperiodic lattices},
  author = {Jiang, H. and Lang, L.-J. and Yang, C. and Zhu, S.-L. and Chen, S.},
  journal = {Phys. Rev. B},
  volume = {100},
  issue = {5},
  pages = {054301},
  numpages = {8},
  year = {2019},
  month = {Aug},
  publisher = {American Physical Society},
  doi = {10.1103/PhysRevB.100.054301},
  url = {https://link.aps.org/doi/10.1103/PhysRevB.100.054301}
}

@article{105,
  title = {Edge States and Topological Invariants of Non-Hermitian Systems},
  author = {Yao, S. and Wang, Z.},
  journal = {Phys. Rev. Lett.},
  volume = {121},
  issue = {8},
  pages = {086803},
  numpages = {8},
  year = {2018},
  month = {Aug},
  publisher = {American Physical Society},
  doi = {10.1103/PhysRevLett.121.086803},
  url = {https://link.aps.org/doi/10.1103/PhysRevLett.121.086803}
}

@article{106,
doi = {10.1088/2399-6528/aab64a},
url = {https://doi.org/10.1088/2399-6528/aab64a},
year = {2018},
month = {Mar},
publisher = {IOP Publishing},
volume = {2},
number = {3},
pages = {035043},
author = {Xiong, Ye},
title = {Why does bulk boundary correspondence fail in some non-hermitian topological models},
journal = {J. Phys. Commun.}
}

@article{107,
  title = {Chaos, Quantum Recurrences, and Anderson Localization},
  author = {Fishman, S. and Grempel, D. R. and Prange, R. E.},
  journal = {Phys. Rev. Lett.},
  volume = {49},
  issue = {8},
  pages = {509--512},
  numpages = {0},
  year = {1982},
  month = {Aug},
  publisher = {American Physical Society},
  doi = {10.1103/PhysRevLett.49.509},
  url = {https://link.aps.org/doi/10.1103/PhysRevLett.49.509}
}

@article{108,
  title = {Quantum dynamics of a nonintegrable system},
  author = {Grempel, D. R. and Prange, R. E. and Fishman, Shmuel},
  journal = {Phys. Rev. A},
  volume = {29},
  issue = {4},
  pages = {1639--1647},
  numpages = {0},
  year = {1984},
  month = {Apr},
  publisher = {American Physical Society},
  doi = {10.1103/PhysRevA.29.1639},
  url = {https://link.aps.org/doi/10.1103/PhysRevA.29.1639}
}

@article{109,
title = {Almost periodic Schrödinger operators IV. The maryland model},
author = {Barry Simon},
journal = {Ann. Phys.},
volume = {159},
number = {1},
pages = {157-183},
year = {1985},
issn = {0003-4916},
doi = {https://doi.org/10.1016/0003-4916(85)90196-4},
url = {https://www.sciencedirect.com/science/article/pii/0003491685901964}
}

@article{110,
title = {Simple models of quantum chaos: Spectrum and eigenfunctions},
author = {Felix M. Izrailev},
journal = {Phys. Rep.},
volume = {196},
number = {5},
pages = {299-392},
year = {1990},
issn = {0370-1573},
doi = {https://doi.org/10.1016/0370-1573(90)90067-C},
url = {https://www.sciencedirect.com/science/article/pii/037015739090067C}
}

@article{111,
  title = {Edge States and Topological Invariants of Non-Hermitian Systems},
  author = {Yao, S. and Wang, Z.},
  journal = {Phys. Rev. Lett.},
  volume = {121},
  issue = {8},
  pages = {086803},
  numpages = {8},
  year = {2018},
  month = {Aug},
  publisher = {American Physical Society},
  doi = {10.1103/PhysRevLett.121.086803},
  url = {https://link.aps.org/doi/10.1103/PhysRevLett.121.086803}
}

@article{112,
  title = {Symmetry and Topology in Non-Hermitian Physics},
  author = {Kawabata, K. and Shiozaki, K. and Ueda, M. and Sato, M.},
  journal = {Phys. Rev. X},
  volume = {9},
  issue = {4},
  pages = {041015},
  numpages = {52},
  year = {2019},
  month = {Oct},
  publisher = {American Physical Society},
  doi = {10.1103/PhysRevX.9.041015},
  url = {https://link.aps.org/doi/10.1103/PhysRevX.9.041015}
}

@article{113,
  title = {Non-Bloch Band Theory of Non-Hermitian Systems},
  author = {Yokomizo, K. and Murakami, S.},
  journal = {Phys. Rev. Lett.},
  volume = {123},
  issue = {6},
  pages = {066404},
  numpages = {6},
  year = {2019},
  month = {Aug},
  publisher = {American Physical Society},
  doi = {10.1103/PhysRevLett.123.066404},
  url = {https://link.aps.org/doi/10.1103/PhysRevLett.123.066404}
}

@article{114,
  title = {Non-Hermitian second-order topological phases and bipolar skin effect in photonic kagome crystals},
  author = {Yang, X. and Feng, Y. and Wahab, A. and Geng, H.},
  journal = {Phys. Rev. A},
  volume = {113},
  issue = {2},
  pages = {023506},
  numpages = {8},
  year = {2026},
  month = {Feb},
  publisher = {American Physical Society},
  doi = {10.1103/s26b-8bdl},
  url = {https://link.aps.org/doi/10.1103/s26b-8bdl}
}

@article{115,
  title = {Non-Hermitian bulk-boundary correspondence in a periodically driven system},
  author = {Cao, Y. and Li, Y. and Yang, X.},
  journal = {Phys. Rev. B},
  volume = {103},
  issue = {7},
  pages = {075126},
  numpages = {10},
  year = {2021},
  month = {Feb},
  publisher = {American Physical Society},
  doi = {10.1103/PhysRevB.103.075126},
  url = {https://link.aps.org/doi/10.1103/PhysRevB.103.075126}
}

@article{116,
  title = {Non-equilibrium dynamics of localization phase transition in the non-Hermitian disorder-Aubry-Andr\'e model},
  author = {Sun, Y.-M. and Wang, X.-Y. and Zhai, L.-J.},
  journal = {Phys. Rev. A},
  volume = {112},
  issue = {2},
  pages = {022204},
  numpages = {13},
  year = {2025},
  month = {Aug},
  publisher = {American Physical Society},
  doi = {10.1103/26fx-j27m},
  url = {https://link.aps.org/doi/10.1103/26fx-j27m}
}

@article{117,
  title = {Impact of nonreciprocal hopping on localization in non-Hermitian quasiperiodic systems},
  author = {Tong, X. and Zhang, Y. and Li, B. and Yang, X.},
  journal = {Phys. Rev. B},
  volume = {111},
  issue = {21},
  pages = {214202},
  numpages = {10},
  year = {2025},
  month = {Jun},
  publisher = {American Physical Society},
  doi = {10.1103/PhysRevB.111.214202},
  url = {https://link.aps.org/doi/10.1103/PhysRevB.111.214202}
}

@article{118,
  title = {Dynamics of a quantum phase transition in the Aubry-Andr\'e-Harper model with $p$-wave superconductivity},
  author = {Tong, X. and Meng, Y.-M. and Jiang, X. and Lee, C. and Neto, G. D. de M. and Xianlong, G.},
  journal = {Phys. Rev. B},
  volume = {103},
  number = {10},
  pages = {104202},
  numpages = {10},
  year = {2021},
  month = {Mar},
  publisher = {American Physical Society},
  doi = {10.1103/PhysRevB.103.104202},
  url = {https://link.aps.org/doi/10.1103/PhysRevB.103.104202}
}

@article{119,
  title = {Dynamical signature of localization-delocalization transition in a one-dimensional incommensurate lattice},
  author = {Yang, C. and Wang, Y. and Wang, P. and Gao, X. and Chen, S.},
  journal = {Phys. Rev. B},
  volume = {95},
  number = {18},
  pages = {184201},
  numpages = {6},
  year = {2017},
  month = {May},
  publisher = {American Physical Society},
  doi = {10.1103/PhysRevB.95.184201}
}

\end{document}